\documentclass{aip-cp}

\usepackage[utf8]{inputenc}
\usepackage[numbers]{natbib}
\usepackage{rotating}
\usepackage{graphicx}
\usepackage{adjustbox}

\begin{document}

% Title portion
\title{An Efficient Test Facility For The Cherenkov Telescope Array FlashCam 
Readout Electronics Production}

\author[aff1]{F. Eisenkolb\corref{cor1}}
\author[aff1]{S. Diebold}
\author[aff1]{C. Kalkuhl}
\author[aff1]{G. Pühlhofer}
\author[aff1]{A. Santangelo}
\author[aff1]{T.~Schanz}
\author[aff1]{C. Tenzer}
\author[aff2]{the FlashCam team}
\author[aff3]{the CTA consortium}

\affil[aff1]{Institut für Astronomie und Astrophysik, 
Kepler Center for Astro and Particle Physics, 
Eberhard Karls Universität Tübingen, Sand 1, 72074 Tübingen, Germany}
\affil[aff2]{Proc. SPIE 9145, Ground-based and Airborne 
Telescopes V, 914531 (22 July 2014)}
\affil[aff3]{See www.cta-observatory.org for full author and affiliation list}
\corresp[cor1]{Corresponding author: eisenkolb@astro.uni-tuebingen.de}

\maketitle

\begin{abstract}
The Cherenkov Telescope Array (CTA) is the planned next-generation 
instrument for ground-based gamma-ray astronomy, currently 
under preparation by a world-wide consortium. 
The FlashCam group is preparing a photomultiplier-based 
camera for the Medium Size Telescopes of CTA, with a fully digital 
Readout System (ROS). For the forthcoming mass
production of a substantial number of cameras, efficient 
test routines for all components are currently under development. 
We report here on a test facility for the ROS components. A test setup 
and routines have been developed and an early version of that setup 
has successfully been used to test a significant fraction of the ROS 
for the FlashCam camera prototype in January 2016. The test setup 
with its components and interface, as well as first results, are 
presented here.
\end{abstract}

\section{INTRODUCTION}

FlashCam for Medium Size Telescopes (MST) is a photo-multiplier tube (PMT) based camera  for 
Cherenkov telescopes. Its main distinctive features are the spatial separation 
between Photon Detection Plane (PDP) and Readout System (ROS), the use of a 
logarithmic amplifier to cover a large input signal range with only one 
channel, and a fully digital ROS. This means that the PMT signal 
is digitized directly after amplification, and that the trigger decision 
is made on digital data. This approach simplifies the analog electronics components
of the camera. In 2016 a prototype camera with about half the number of 
pixels of a full FlashCam MST camera 
has been built and it is currently under severe testing at the Max-Planck-Institute for
Nuclear Physics (MPIK) institute in Heidelberg. The current schedule foresees the fabrication of 
two pre-production cameras until 2018. To guarantee the highest reliability,
all components of the ROS electronics (as well as all other components)
will be tested,
and their performance and functionality verified at the Institute for Astronomy
and Astrophysics Tübingen (IAAT). To this end,
a test facility has been designed whose devices and procedures were already used early in 
2016 to test most of the ROS electronics for the FlashCam prototype and some more improvements 
have also been implemented for the application during pre-production and production.

\begin{center}
\begin{table}[ht]
\label{tab:a}
\begin{adjustbox}{width=\textwidth}
\begin{tabular}{c c}
  \includegraphics[width=0.6\textwidth]{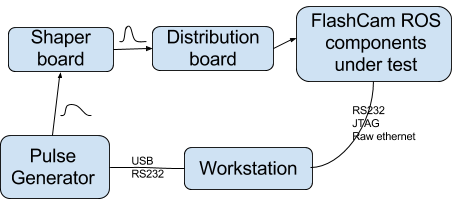}
  &
  \includegraphics[width=0.3\textwidth]{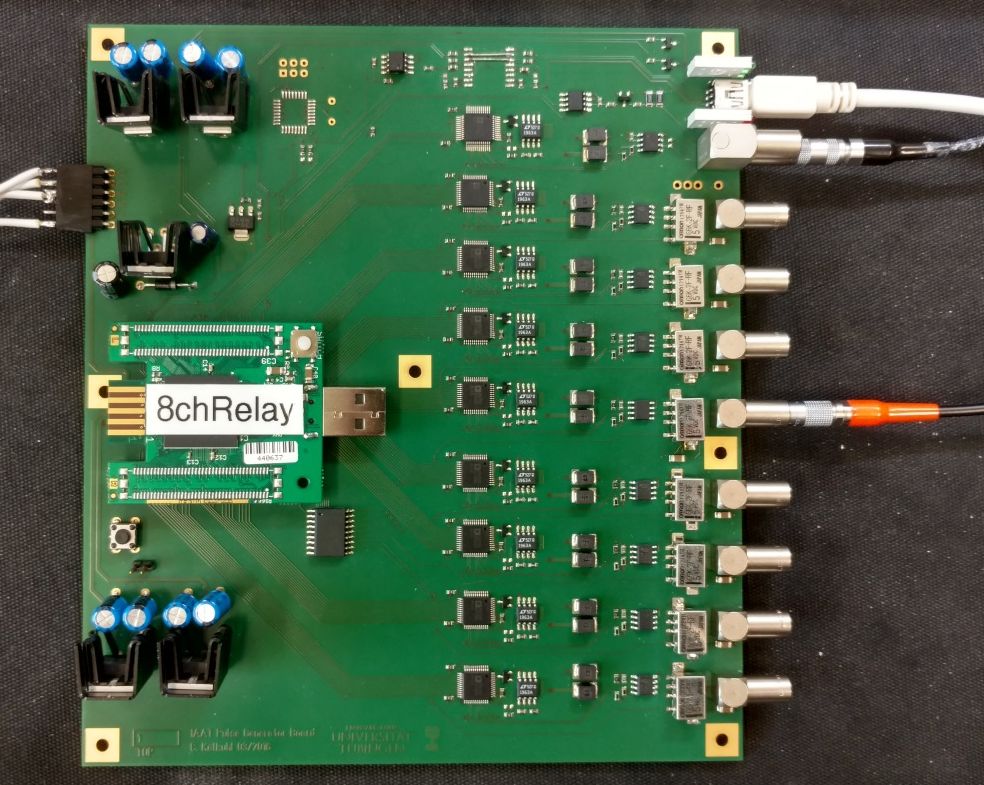} \\
  \textbf{FIGURE 1.} Sketch of the FlashCam ROS test-facility components.
  & \textbf{FIGURE 2.} The pulse generator designed at IAAT.
\end{tabular}
\end{adjustbox}
\end{table}
\end{center}

\section{TEST FACILITY}
% * <feisenkolb@gmail.com> 2016-09-13T07:32:08.547Z:
% 
% - More details on the signal (capacitor)
% - Signal evaluation
% - What is evaluated automatically
% - Graphics for the connections
% - Image of preliminary lab space?
% 
% ^.

The idea of the test facility is to quickly evaluate the functionality of 
the ROS electronics, independently 
of  the  Photon Detection Plane. The ROS consists of multi-purpose motherboards
which can host one of three different kinds of daughter-boards.
The first option is to mount two FADC daughter boards
\footnote{Flash Analog To Digital Converter. Daughter boards equipped 
with FADC chips to sample data at 250 MHz.} 
for data sampling (up to 24 channels per motherboard).
The second option is to mount one trigger daughter board to synchronize timing 
and distribute data between motherboards with FADC daughter boards 
(up to eight FADC equipped boards per trigger card). 
The third and last option is a master daughter board
to do camera wide synchronization and trigger distribution and to interact with the
telescope array. 

The general layout of the facility 
is shown in Figure 1. Being able to evaluate the ROS independently of the
PDP minimizes the complexity of the test environment. On one side, it minimizes
the components involved, on the other side, it provides a known and easily controllable
input to the ROS.
To achieve this, the PDP is replaced by a combination of a pulse 
generator 
and a distribution/shaper board, which shapes the 
signal emitted by the pulse generator to closely resemble that of the FlashCam PDP. 
The pulse generator which was developed at IAAT is shown in Figure 2. 
It features eight channels with a maximum output amplitude of 3.5 V and a resolution 
of 10 bits. The rise time of two nanoseconds is critical for the shaped pulse
to closely resemble that of the original PDP. To cover the whole signal range
with sufficient accuracy an additional 20 dB attenuator can be switched in.
The pulse generator can be interfaced via USB and an alternative serial 
connection via RS232 is currently being implemented. The goal of this
development is to be independent of any software or drivers
installed on lab PCs. The pulse generator provides a trigger output. The possibility
to trigger the pulse generator externally is currently under development.
A workstation running CentOS is
used to control the pulse generator and it also works as an interface to the 
FlashCam ROS. The pulses emitted by the pulse generator, sketched in Figure 1, are sent to a pair of distribution and 
shaper boards developed at MPIK.

The basic idea is to have a signal with 
a very short rise time (smaller than several nanoseconds)
and differentiate it. The resulting signal will have an approximate 
duration of that of the rising edge of the original signal and resemble 
in shape that of a PMT. 

This PMT-like pulse is then shaped by an electronics 
circuit that resembles that of the preamplifier in the PDP of FlashCam.
Both differentiation and shaping are done by the shaper board.
The shaped pulses are then sent to FlashCam FADC boards, 
where they are digitized and stored in a ring buffer. This ring buffer
can then be read out via a microblaze microcontroller inside the Spartan 6
FPGA on board of the FlashCam motherboards and it is then transmitted to
the lab workstation. 

All programmable options, such as the ADC baseline,
can be set and tested, as well as signals with different amplitudes 
can be applied, read out and evaluated to cover the whole dynamic range. 
Higher level parameters such as the linearity of the reconstructed 
amplitude and the stability of the timing are analyzed automatically.

The measured data is then analyzed. First of all concerning the amplitudes a linear behavior is
expected, thus a linear fit is applied and a verification on the fit parameters is applied to
check if they are inside
the specifications. For data that is not analytic, such as the full width at half maximum 
of the original pulse or 
center of gravity of the time derivative of the signal (which is used for timing), 
min/max curves are defined and it is checked whether the data falls into the defined regime.

More detailed electronics tests are only
performed if the functionality tests fail. Then the boards are closely evaluated to
find the cause of error, to give precise feedback to the manufacturer, and, if possible,
fix the electronics in house. 

The FlashCam ROS has two interface options. On a single board level connections can 
be made via RS232 interface or via raw Ethernet, which is a custom low level 
Ethernet protocol.
The serial connection can be used to verify low level functionality whereas
the raw Ethernet connection is used for normal operation and data transfer.
Control of the test environment can happen either very directly via command line
tools, alternatively a GUI gives a higher level interface and automates 
most of the test control.

The procedures for  temperature and vibration tests
is currently under evaluation. These are particularly interesting because they
show dormant faults that may only occur later in a products lifetime, such as
cold solders. Currently we plan to do temperature cycling
over the storage temperature range for powered off devices as well as cycling over the
operational range with powered devices. Exact specifications for these tests are 
currently being defined. Temperature cycling is favored over dedicated burn-in at the
module level (see \citet[chap. ~7]{bib3} for details).

\section{RESULTS FROM FIRST PROTOTYPE TESTS AND OUTLOOK}
% * <feisenkolb@gmail.com> 2016-09-13T07:36:02.397Z:
% 
% Images
% 
% ^.
In January, a first version of the setup has been used at MPIK to test almost the complete 
ROS electronics for the FlashCam prototype. This included 104 motherboards and 
140 FADC daughterboards, which corresponds to 1680 channels. 

%TODO more abstract
The tasks performed included the final assembly of the motherboards, optical inspection,
configuration of the single boards as well as running the complete test suite.
This verified the correct functionality of the ROS components, including configuration
and event reconstruction with the required performance.
Only a small number of components showed faults that were already identified in the
first steps of configuration.
Besides functionality and quality control and verification this test provided valuable
input for the planned automation of evaluation of the test results.
Figure~\ref{figsigma} shows an example result. It shows the spread of the amplitudes at different
input amplitudes over all motherboards. This has been done without cross calibration between the
different boards. This shows again the functionality and stability of the FlashCam concept. The 
automated evaluation and control of the tests has since then been implemented.
Currently the exact failure conditions and tolerances are being specified and the test facility 
is being prepared for future test runs.
The inclusion of temperature cycling and vibration tests is being evaluated and
the requirements for temperature cycling are finalized.

\setcounter{figure}{2}
\begin{figure}
\centering
  \includegraphics[width=.7\textwidth]{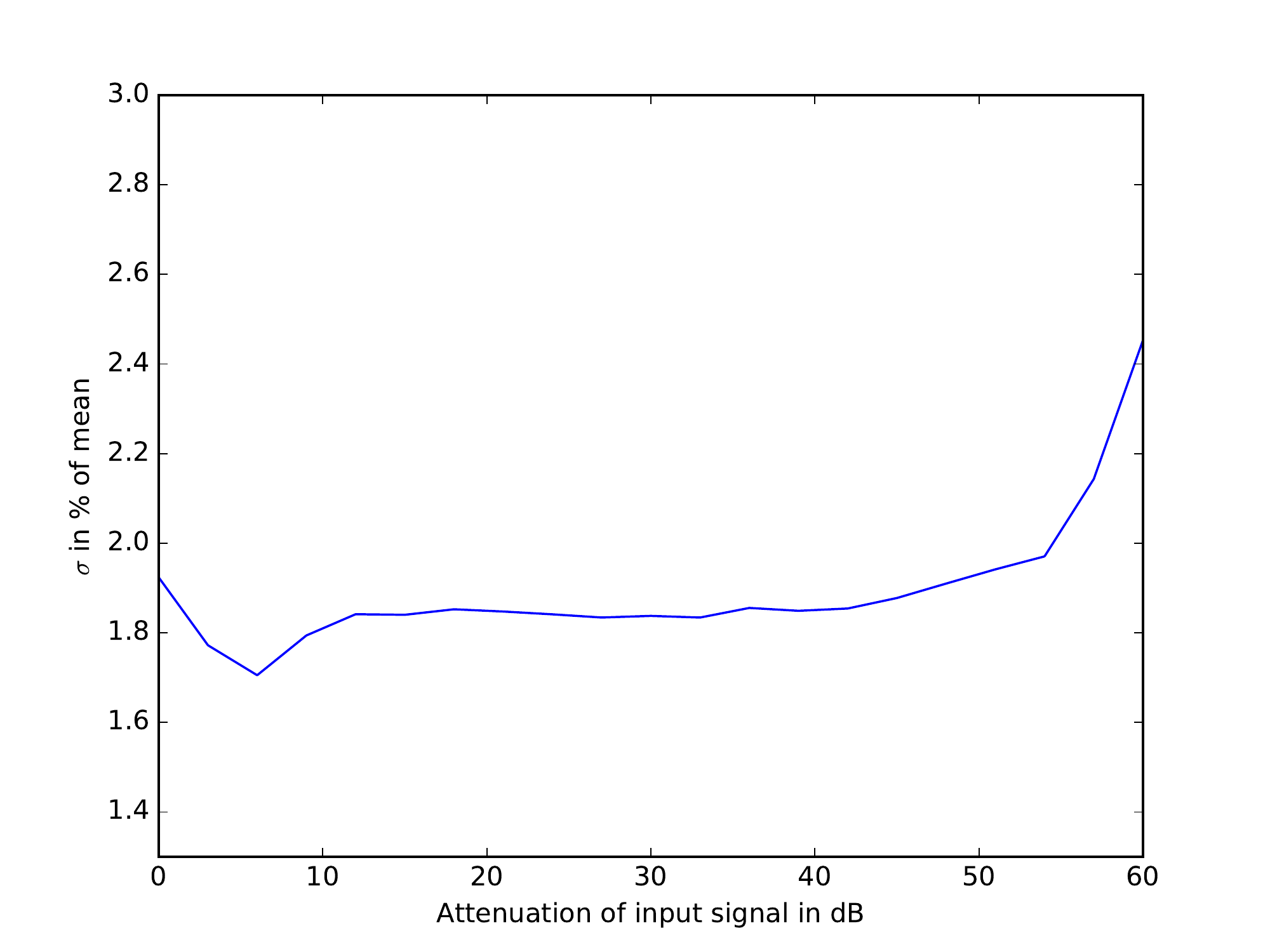}
  \caption{Spread of the reconstructed amplitude over all channels over all motherboards. Even though there was no cross calibration, the amplitude spread was small and again verified the functionality and stability of the FlashCam concept.}
  \label{figsigma}
\end{figure}

% Acknowledgement

\section{ACKNOWLEDGMENTS}
We gratefully acknowledge support from the agencies and organizations under Funding Agencies at 
www.cta-observatory.org

% References

\nocite{*}
\bibliographystyle{aipnum-cp}%
\bibliography{FCROtests}%

\end{document}